\documentstyle[preprint,aps,multicol,epsfig]{revtex}
\tightenlines
\begin{document}

\title{Quantum Critical Point 
in Electron-Doped Cuprates}

\author{C. Kusko$^1$, R.S. Markiewicz$^1$, M. Lindroos$^{1,2}$, and A. 
Bansil$^1$}
\address{1. Physics Department, Northeastern University, Boston MA 02115, USA; 
2. Institute of Physics, Tampere University of Technology, 33101 Tampere, 
Finland}
\maketitle
\begin{abstract}

We analyze doping dependent spectral intensities and Fermi surface maps 
obtained recently in Nd$_{2-x}$Ce$_x$CuO$_{4\pm\delta}$ (NCCO) 
via high resolution 
ARPES measurements, and show that 
the behavior of this electron-doped compound can be 
understood as the closing of a Mott
(pseudo) gap, leading to a quantum critical point just above optimal 
doping. The doping dependence of the effective Hubbard $U$ adduced by 
comparing theoretical and experimental spectra is in resonable 
accord with various estimates and a simple screening calculation.

\end{abstract}
\narrowtext

Angle-resolved photoemission spectra (ARPES) follow a remarkably different 
route with doping in the {\it hole-doped} La$_{2-x}$Sr$_x$CuO$_4$ 
(LSCO)\cite{ZZX} compared to the {\it electron-doped} 
NCCO\cite{nparm}.  Starting from a Mott insulator near $x=0$, LSCO
exhibits the appearance of dynamic or static stripes\cite{ZZX}, 
with a concomittant discontinuity in the chemical potential\cite{AIP}.  
In sharp contrast, in NCCO, characteristic signatures of stripe order, e.g.
the 1/8 anomaly and the NQR wipeout, appear greatly attenuated if not 
absent\cite{AKIS}, and the Fermi level shifts smoothly into the upper Hubbard 
band\cite{AIP}.  Here we show how the salient features of the most recent 
high resolution ARPES data in NCCO\cite{nparm} can be understood in terms of the
behavior of a uniformly doped Mott insulator. The doping dependence of the 
effective Hubbard $U$ parameter adduced from the experimental spectra is in 
reasonable accord with various estimates, including a computation in which 
we screen the bare $U$ via interband excitations. A quantum critical 
point (QCP) where the Mott gap closes is predicted just beyond optimal doping.
Finally, we have carried out extensive first-principles simulations of 
photointensities in order to ascertain that the key spectral features 
discussed in this article (e.g. the appearance and growth of the 
$(\pi/2,\pi/2)$ centered hole orbit with doping) are genuine effects of 
electron correlations beyond the conventional LDA framework and not 
related to the energy and k-dependencies of the ARPES matrix 
element\cite{matri}. 

Our analysis of the doping dependence of the Mott gap -- perhaps better 
referred to as a pseudogap\cite{MW,mw}, invokes a mean field solution to the 
one band\cite{3b,MK2} Hubbard model. The mean field approach provides  
a good description of not only the pseudogap in the 1D 
charge density wave (CDW) systems\cite{LRA},
but also of the undoped insulator in the 2D Hubbard problem, including the 
presence of the `remnant Fermi surface' (rFS)\cite{rfs} seen in 
ARPES experiments\cite{CCOC} and Monte Carlo simulations\cite{BSW}; 
the spin density wave (SDW) excitations are described by fluctuations 
about the mean field\cite{SWZ}.  The mean field theory correctly predicts
a (stripe) phase instability (negative compressibility) associated with
hole doping\cite{SiTe}; whereas assuming a finite second-neighbor hopping 
$t'<0$, for electron doping the compressibility is positive\cite{MK1}, 
suggesting that a uniformly doped AFM state should be stable.  The preceding 
considerations argue that the mean field would provide a reasonable 
model for discussing the doping dependence of the 
pseudogap. This viewpoint is further supported by mode-coupling 
calculations\cite{MK2}, which have been applied previously to approximately
describe stripes in terms of a CDW in hole-doped cuprates\cite{Cast2,RM5}. 

Concerning methodology, we specifically consider the one-band Hubbard 
model where the Mott gap $\Delta$ is well known to arise (in the mean 
field SDW formulation\cite{SWZ}) from a finite expectation value of the 
magnetization $m_{\vec Q}$ at the wave vector $\vec Q=(\pi,\pi)$.  The 
self-consistent gap equation is 
\begin{equation}
1=U\sum_k {f(E^{v}_{\vec k})-f(E^{c}_{\vec k})\over E_{0\vec k}},
\label{eq:1}
\end{equation}
where $f$ is the Fermi function and  
\begin{equation}
E^{c,v}_{\vec k}={1\over 2}(\epsilon_{\vec k}+\epsilon_{\vec k+\vec q}\pm E_
{0\vec k})
\label{eq:2}
\end{equation}
Here, the superscript $c$ refers to the upper Hubbard band (UHB) 
and goes with the + sign on the right side while $v$ goes with the 
- sign and the lower Hubbard band (LHB)\cite{ZR1}.
$E_{0\vec k}=\sqrt{(\epsilon_{\vec k}-\epsilon_{\vec k+q})^2+4\Delta^2}$, and 
the independent particle dispersion is: $\epsilon_k=-2t(c_x+c_y)-4t'c_xc_y$, 
$c_i=\cos{k_ia}$.  We study Eq.~\ref{eq:1} self-consistently as a function of 
electron doping, assuming that the mean-field transition temperature corresponds
to the experimentally observed pseudogap. 
$U$ is thus treated as an effective parameter $U_{eff}$ to fit the experimental 
data.  The Green's function in the antiferromagnetic (AFM) ground state is 
given by
\begin{equation}
G(\vec {k},\omega)={u^2_{\vec {k}}\over {\omega-E^{c}_{\vec {k}}}}+{v^2_{\vec 
{k}}\over {\omega-E^{v}_{\vec {k}}}}
\label{eq:3}
\end{equation}
where 
\begin{equation}
u^2_{\vec k}={1\over 2} 
(1+{\epsilon_{\vec k}-\epsilon_{\vec k+\vec Q}\over 2E_{0\vec k}}), 
v^2_{\vec k}={1\over 2} (1-{\epsilon_{\vec k}-\epsilon_{\vec k+\vec Q}
\over 2E_{0\vec k}})
\label{eq:4}
\end{equation}
are the coherence factors. 

Using Eqs. 1-4, we discuss spectral density $A(\vec k, E)$ (given by 
the imaginary part of the Green's function) and the related energy 
dispersions for various doping levels. The FS maps are obtained by taking 
appropriate cuts through $A(\vec k, E)$. Although such theoretical 
maps do not account for the effects of the 
ARPES matrix element\cite{matri}, 
these maps are relevant nevertheless in gaining 
a handle on the FS topology expected in the correlated 
system from ARPES experiments. In any event, we have also carried out 
first-principles simulations based on the 
conventional LDA picture where the ARPES matrix element is properly 
treated and the photoemission process is modeled including full 
crystal wavefunctions in the presence of the surface -- see
Refs.~\onlinecite{matri,LSB} for details of the methodology.

Fig.~\ref{fig:1} displays the doping dependence of the spectral weight in the
vicinity of the $(\pi/2, \pi/2)$ and $(\pi ,0)$ points. 
Fig.~\ref{fig:1}(a) shows that around 
$(\pi/2, \pi/2)$, with increasing doping, the spectral weight shifts rapidly 
towards the $E_F$ as the gap between the LHB and UHB decreases.
At $x=0.15$, the gap is quite small and the LHB and UHB overlap. 
In contrast, in the momentum region near 
$(\pi,0)$, Fig.~\ref{fig:1}(b), both LHB and UHB are more directly 
involved. Even at small doping levels ($x=0.04$), $E_F$ intersects the 
bottom of the UHB. With increasing electron concentration, the UHB moves 
to lower energies while the LHB shifts closer to the $E_F$ as the gap 
decreases. These results are remarkably consistent with the corresponding 
ARPES data; Fig. 2a of Ref.~\onlinecite{nparm} shows a rapid movement of 
spectral weight at $(\pi /2,\pi /2)$ from $~1.3$ eV to around 0.3 eV binding 
energy close to $E_F$.  This effect, although somewhat 
less clear, is seen at $(\pi,0)$ 
as well (Fig. 2b of Ref.~\onlinecite{nparm}): the weight near $E_F$ grows 
faster with doping, but the LHB is at a lower binding energy, and is less 
clearly resolved from the background.  
This doping dependence suggests that electrons first enter the UHB near 
$(\pi,0)$.

The energy dispersions of Fig. 2 are useful not only in understanding the 
spectra of Fig. 1, but also provide a handle on the FS maps expected in 
ARPES experiments. The $E_F$ is seen in Fig. 2 to rise smoothly with respect to 
the UHB with increasing electron doping, whereas in the hole doped 
cuprates, the $E_F$ gets pinned by the stripes near the mid-gap region over a 
large doping range. We should keep in mind that various bands do not possess 
the same spectral weight -- this point is emphasized by depicting the 
coherence factors of Eq. 4 via the width of the bands in Fig. 2. As the 
Mott gap nearly collapses, the thick lines in Fig. 2(d) essentially present 
the appearance of the uncorrelated band structure. 
   
Fig. 3 illustrates the evolution of the FS corresponding to the band structure 
of Fig. 2. For low $x$, $E_F$ lies near the bottom of the UHB and gives rise to
electron pockets centered at $(\pi,0)$ and $(0,\pi)$ points in 
Fig.~\ref{fig:3}(a). The shadow segments of the FS are clearly 
evident in Fig.~\ref{fig:3}(b) where half of the electron pocket gains
spectral weight at the expense of the other half due to the decreasing 
intensity in the magnetic Brillouin Zone (BZ) via the coherence 
factors; a weak imprint of the magnetic BZ boundary (the diagonal
line connecting $(0,\pi)$ and $(\pi,0)$ points) can also be seen. 
By $x=0.15$, the gap shrinks considerably, and the $E_F$ begins intersecting 
the LHB around $(\pi/2,\pi/2)$. The FS now consists of three sheets: 
electron-like sheets near $(\pi,0)$ and $(0,\pi)$ and a hole-like sheet around 
$(\pi/2,\pi/2))$ separated by a residual gap located at the 
`hot-spots'\cite{HlR} along the BZ diagonal. Interestingly, in this doping 
regime, transport studies find evidence for two band conduction, and a change in
the sign of the Hall coefficient\cite{nctran}; the hole pocket associated with 
the LHB can explain both these effects.  

We stress that the FS evolves following a very 
different route in the hole-doped case for the $t-t'$ one band Hubbard 
model\cite{ChubMorr} (neglecting stripes). At low doping levels, the FS 
consists of small hole pockets around $(\pi/2,\pi/2)$ points. With 
increasing hole concentration, these pockets increase in size and 
merge to yield a large $(\pi,\pi)$ centered FS 
satisfying the volume constraints of the Luttinger theorem.

Figure 4 directly compares the theoretical FS maps against the corresponding 
experimental results. Here we have included a small second neighbor hopping 
parameter $t''=0.1t$ in the computations in order to account for the slight 
shift of the center of the hole pocket away from $(\pi/2,\pi/2)$ in the 
experimental data, even though the maps of Fig. 3 with only $t-t'$ already 
provide a good overall description (after resolution broadening) of the 
measurements. The agreement between theory and experiment in Fig. 4 is 
remarkable: Both sets of maps show the electron pockets at low 
doping ($x=0.04$); the beginning of the hole pocket at $x=0.10$, 
together with the 
presence of shadow features around electron pockets and the appearance of 
intensity around the magnetic BZ; and finally, at $x=0.15$, the evidence 
of a well formed hole pocket and a three-sheeted FS which has begun to 
resemble the shape of a large $(\pi,\pi)$ centered hole sheet. Fig. 4 
displays discrepancies as well (e.g. the hole pocket related feature 
appears double peaked in experiments, but not in theory, and other 
details of spectra around the $E_F$), but this is not surprising since 
we are invoking a rather simple one-band Hubbard model and the effects of
the ARPES matrix element are missing in these calculations. 

The parameters used in the computations of Fig. 4 are as follows.\cite{ueff}
For the half-filled case, the values are identical to those used 
previously\cite{OSP}: $t=0.326 eV$, $t'/t=-0.276$, $U=6t$. 
For finite doping $x$, the only change is that $U$ is 
assumed to be $x$-dependent, $U=U_{eff}(x)$, and $t''=0.1t$. 
The actual values of $U_{eff}$ used (solid dots in 
Fig. 5(a)), are in reasonable accord with various estimates shown in Fig. 5(a) 
which are: the approximate value from Kanamori (arrow) for a nearly empty 
band\cite{Kana}, and Monte Carlo results\cite{BSW2,ChAT}, for $t'=0$ (star). 
Finally, we have carried out a computation of {\it screened} $U$ using 
$U_{eff}=U/(1+<P>U)$, taking $P$ as the charge susceptibility which includes 
only interband contributions with bare $U=6.75t$. The $U_{eff}$ so obtained 
for electrons (solid line) and holes (dashed) is shown\cite{Chen0}. 

Fig. 5(b) considers the behavior of the staggered 
magnetization, $m_{\vec Q}(x)$, 
for electrons and holes using the computed values of $U_{eff}$ for 
{\it holes} given by dashed line in Fig. 5(a).\cite{calc} We see that 
$m_{\vec Q}(x)$ and hence the pseuodgap, which is proportional to $m_{\vec Q}(x)
$, vanishes slightly above optimal doping, yielding a QCP. Notably, 
superconductivity near an AFM QCP has been reported in a number of 
systems\cite{MaLo}. However, the present case is different in that we have a 
mean-field QCP associated directly with short-range AFM fluctuations and the 
{\it closing of the Mott gap}.  Since there is no interfering phase 
separation instability, the bulk N\'eel transition persists out to 
comparable, but clearly lower dopings, with $T_N\rightarrow 0$ near 
$x=0.13$. 

We have carried out extensive first-principles simulations of the ARPES 
intensities in NCCO within the LDA framework 
for different photon energies, polarizations and surface 
terminations, in order to ascertain the extent to which the characteristics 
of measured FS maps could be confused with the effects of the ARPES matrix 
element\cite{matri} missing in the computations of Figs. 3 and 4. The computed 
FS maps including full crystal wavefunctions of the initial and final states 
in the presence of the Nd-CuO$_2$-Nd-O$_2$-terminated surface at 16 eV are 
shown in Fig. 6 for two different polarizations, and are typical. 
The intensity is seen 
to undergo large variations as one goes around the $(\pi,\pi)$ centered 
hole orbit, and to nearly vanish along certain high symmetry lines in some 
cases. Nevertheless, we do not find any situation which resembles
the doping dependencies displayed in Figs. 3 and 4. It is clear that strong
correlation effects beyond the conventional LDA-based picture are needed
to describe the experimental ARPES spectra and that the $t-t'-t''$ 
Hubbard model captures some of the essential underlying physics. 

In conclusion, this study indicates that the electron doped NCCO is an ideal
test case for investigating how superconductivity arises near a 
QCP in doped Mott insulators, untroubled by 
complications of stripe phases. The doping dependence of $U_{eff}$
adduced in this work has implications in understanding the behavior 
of the cuprates more generally since the pseudogap in both the electron
and the hole doped systems must arise from the same Mott gap at 
sufficiently low doping. 

We thank N.P. Armitage and Z.-X. Shen for sharing their data with us 
prior to publication. This work is supported by the U.S.D.O.E. 
Contract W-31-109-ENG-38 and benefited from the allocation of 
supercomputer time at the NERSC and the Northeastern University 
Advanced Scientific Computation Center (NU-ASCC).

\begin{figure}
\vskip0.5cm 
\caption{Integrated spectral weight is shown near (a): $(\pi /2,\pi /2)$, 
and (b): $(\pi ,0)$ points, for four different doping levels, $x=$ 0.0-0.15, 
as marked in (b). Domains in the Brillouin zone over which the spectral 
weight was integrated are shown in the insets. Fermi energy defines the 
energy zero in all cases.}
\label{fig:1}
\end{figure}
\begin{figure}
\vskip0.5cm 
\caption{Energy dispersions for various doping levels $x$. Energy zero defines
the Fermi energy as in Fig. 1. Thickness of lines represents the spectral 
weights of various bands given by the coherence factors of Eq. 4.}
\label{fig:2}
\end{figure}
\begin{figure}
\vskip0.5cm 
\caption{Fermi surfaces corresponding to the band structures of Fig. 2. 
Maps are obtained by integrating the spectral density function 
(proportional to $Im[G]$, Eq. 3) over an energy window of 60 meV around 
$E_F$; highs denoted by red and lows by blue. Experimental resolution 
effects are not included for the illustrative purpose of this figure.}
\label{fig:3}
\end{figure}
\begin{figure}
\vskip0.5cm 
\caption{Upper panels: Theoretical Fermi surface maps including 
resolution broadening for the $t-t'-t''$ model (see text) for
different doping levels $x$. 
Lower panels: Corresponding experimental maps after 
Ref. \protect\onlinecite{nparm}. Color scheme as in Fig. 3. }
\label{fig:3a}
\end{figure}
\begin{figure}
\vskip0.5cm 
\caption{(a): $U_{eff}$ (scaled by nearest neighbor hopping parameter $t$), 
and (b): staggered magnetization $m_{\vec Q}(x)$, vs doping $x$ for 
electrons (solid lines) and holes (dashed lines). Filled circles give 
values used in the computations of Fig. 4 and are representative of 
the experimental ARPES data in NCCO. Values of $U_{eff}$ given by 
Kanamori\protect\onlinecite{Kana} (arrow) and Monte-Carlo 
studies\protect\onlinecite{BSW2,ChAT} (star) are shown in (a). }
\label{fig:5}
\end{figure}
\begin{figure}
\vskip0.5cm 
\caption{ Theoretical FS maps in NCCO obtained via first principles 
simulations which include the effect of the ARPES matrix element but not
of strong correlations for two different polarizations (given by the white 
arrows) at a photon energy of 16 eV. Color scheme as in Fig. 3. } 
\label{fig:4}
\end{figure}

\begin{references}
\bibitem{MW}Recalling Mermin-Wagner theorem \protect\onlinecite{mw}, 
two dimensional fluctuations suppress long range order, 
although a real gap (long-range N\'eel order) can appear at low temperatures 
due to residual interlayer coupling.
\bibitem{ueff}In Figs. 1-3, where $t''=0$, $U_{eff}/t$ values 
used are virtually the same as
those given by filled dots in Fig. 5(a) for $x=$0 and 0.04; for $x=$0.10 and 
0.15, the values used are $\approx$ 15$\%$ smaller.
\bibitem{calc}This procedure is not self-consistent for electrons, but merely 
provides a smooth $U_{eff}(x)$ curve matching the solid dots 
determined from experiment.
\end{references}
\end{document}